\documentclass[12pt]{article}

\usepackage{amsmath,amssymb,amsfonts}
\usepackage{graphicx}

\makeatletter

\renewcommand\section{{\setcounter{equation}{0}}
			       \@startsection{section}{1}{\z@}
                                   {-3.5ex \@plus -1ex \@minus -.2ex}
                                   {2.3ex \@plus .2ex}
                                   {\normalfont\large\bfseries}}
\renewcommand\subsection{\@startsection{subsection}{2}{\z@}
                                   {-3.25ex\@plus -1ex \@minus -.2ex}
                                   {1.5ex \@plus .2ex}
                                   {\normalfont\normalsize\bfseries}}
\renewcommand\subsubsection{\@startsection{subsubsection}{3}{\z@}
                                   {-3.25ex\@plus -1ex \@minus -.2ex}
                                   {1.5ex \@plus .2ex}
                                   {\normalfont\normalsize\bfseries}}
\renewcommand\paragraph{\@startsection{paragraph}{4}{\z@}
                                   {3.25ex \@plus1ex \@minus.2ex}
                                   {-1em}
                                   {\normalfont\normalsize\bfseries}}

\makeatother

\newcommand{\beq}{\begin{equation}}
\newcommand{\eeq}{\end{equation}}
\newcommand{\bea}{\begin{eqnarray}}
\newcommand{\eea}{\end{eqnarray}}

\newcommand{\U}{{\rm U}}

\newcommand{\USp}{{\rm USp}}
\newcommand{\Spin}{{\rm Spin}}

\newcommand{\id}{\hbox{1\kern-.27em l}}

\newcommand{\cF}{{\cal F}}

\newcommand{\cN}{{\cal N}}
\newcommand{\RR}{\mathbb{R}}

\renewcommand{\d}{\partial}
\newcommand{\iso}{\cong}

\long\def\symbolfootnote[#1]#2{\begingroup \def\thefootnote{\fnsymbol{footnote}}\footnote[#1]{#2}\endgroup} 

\begin{document}

\pagestyle{empty}

\begin{center}
\vspace*{30mm}
{\LARGE $(2,0)$ theory on Taub-NUT: A note on\\WZW models on singular fibrations}

\vspace*{20mm}
{\large
Fredrik Ohlsson\symbolfootnote[1]{fredrik.ohlsson@chalmers.se}
}

\vspace*{5mm}
Department of Fundamental Physics\\
Chalmers University of Technology\\
S-412 96 G\"oteborg, Sweden

\vspace*{30mm}{\bf Abstract:}
\end{center}
In this note we consider the gauge field equation of motion for the dimensional reduction of the $(2,0)$ tensor multiplet on singular circle fibrations. The fibrations are characterized by the corresponding $\U(1)$ action having a codimension four fixed point locus $W$. Along $W$, the dimensional reduction of the $(2,0)$ receives a modification described by a WZW model. We consider the emergence of the additional degrees of freedom through the topological term in the action, which in addition to the gauge field strength involves the $\U(1)$ connection of the space-time fibration. We also consider the Taub-NUT space as a simple example of a singular fibration, and in particular consider spherically symmetric solutions for the field strength.

\newpage
\pagestyle{plain}

\hrule
\tableofcontents
\vspace*{6mm}
\hrule

\section{Introduction}
In this note we continue the investigation initiated in~\cite{Linander:2012} of the dimensional reduction of the tensor multiplet of $(2,0)$ theory on a (Lorentzian) manifold $M_6$ which is a six-dimensional circle fibration. More precisely, following the general results outlined in~\cite{Witten:2009b} we will consider a concrete example of a singular fibration. A regular $\U(1)$ fibration $M_6 \to M_5$ can be described by a free action of $\U(1)$ on the manifold $M_6$, in which case the projection to the base space is simply the quotient map to $M_5 = M_6 / \U(1)$. For a singular fibration we relax the condition that the $\U(1)$ action be free and allow it to have fixed points. In particular, we will consider the case where the action of $\U(1)$ has a two-dimensional fixed point locus $W$ and the quotient $M_5 = M_6 / \U(1)$ is in fact a smooth manifold\footnote{A detailed discussion of the properties of this construction can be found in~\cite{Witten:2009b}.}. However, due to the presence of fixed points the description of $M_6$ as a $\U(1$) bundle is only valid over $M_5 \! \setminus \! W$.

The low-energy effective action for the tensor multiplet dimensionally reduced on a $S^1$ fibration is (abelian) supersymmetric Yang-Mills theory which in particular contains a topological term $\int \theta \wedge F \wedge F$ for the gauge field. The explicit form of the action for this theory was obtained in~\cite{Linander:2012} and will be reviewed in part below. For the case of a codimension four fixed point singularity it is shown in~\cite{Witten:2009b} that the anomalous transformation of the topological term under gauge transformations that are non-trivial along $W$ requires the introduction of a gauged WZW model~\cite{Witten:1984,Witten:1992} supported on $W$. (For $M_5 \! \setminus \! W$ the low-energy theory receives no modification.)

The purpose of the present note is to consider in some detail a simple example of the singular situation described above, and in particular the degrees of freedom localized at $W$ and their coupling to the Yang-Mills theory. We will therefore take $M_6 = \RR^{1,1} \times TN$, where $TN$ is the Taub-NUT space, with a product metric. The action of $\U(1)$ is trivial in the first factor and described below for the second one. Under the assumption of spherical symmetry in the $\RR^3$ of $TN$, we consider the solution of the equations of motion for $F$ and discuss the coupling to the current of the WZW model induced by the singularity $W$.

The outline of the paper is the following: In section 2 we review certain results of~\cite{Linander:2012}, valid away from $W$, and discuss implications of the singularity for general geometries. We then consider in section 3 the specialization to the Taub-NUT geometry and the fixed point locus of the $\U(1)$ action on $M_6$ of the particular form above. The corresponding obstruction to extending the $\U(1)$ bundle over all of $M_5$ and the induced modification along $W$ are also discussed. Section 4 is then concerned with the computation of the spherically symmetric solutions to the gauge field strength equations of motion.

\section{$(2,0)$ theory on circle fibrations}
In this section we recall some results obtained in~\cite{Linander:2012,Witten:2009b} for the dimensional reduction of the free tensor multiplet of $(2,0)$ on a regular circle fibration $M_6 \to M_5$ with metric
\beq
ds^2 = g_{\mu \nu} dx^{\mu} dx^{\nu} + R^2 (\varphi + \theta_{\mu} dx^{\mu})^2 \,.
\eeq
Here $x^{\mu}$, with $\mu = 0,1,\ldots,4$, parametrize $M_5$, $g_{\mu \nu}$ is the base space metric, $R$ is the radius of the $S^1$ fibre and $\theta$ is the connection of the $\U(1)$ bundle with curvature $\cF = d\theta$. The low-energy effective theory obtained by dimensional reduction on $S^1$ is supersymmetric Yang-Mills theory. The field strength $F$ is obtained by decomposing the three-form of the tensor multiplet as
\beq
\label{eqn:DecompositionH}
H = E + F \wedge d \varphi
\eeq
and using the self-duality of $H$ to eliminate $E$ in favor of $F$, which then satisfies
\beq
dF = 0
\eeq
\beq
\label{eqn:YangMillsEquationOfMotion}
-d\left( \frac{1}{R} \ast_g F \right) + \cF \wedge F = 0 \,.
\eeq
The first equation is the Bianchi identity of the $F = dA$ gauge field strength and the second is the equation of motion, which can be integrated to an action containing the topological term mentioned in the introduction
\beq
\label{eqn:YangMillsAction}
S_{\rm{YM}} = \int_{M_5} \left( -\frac{1}{R} F \wedge \ast_g F + \theta \wedge F \wedge F \right) \,.
\eeq

The dimensional reduction of the remaining fields of the tensor multiplet produces on $M_5$ a chiral Lorentz spinor $\psi$ and a scalar $\phi$ transforming respectively in the $\mathbf{4}$ spinor representation and the $\mathbf{5}$ vector representation of the $\USp(4) \iso \Spin(5)$ R-symmetry group. Together with the gauge field they comprise an $\cN=4$ abelian vector multiplet on $M_5$, whose dynamics and supersymmetry transformations, derived in~\cite{Linander:2012}, are modified by the fibration geometry.

Relaxing the condition that $M_6$ is a $\U(1)$ bundle over all of $M_5$, by allowing a fixed point locus $W$ as described in the introduction, implies that the curvature $\cF$ develops a singularity along $W$ such that
\beq
\label{eqn:CurvatureSingularity}
d \cF = c_{\cF} \delta_W \,,
\eeq
where $\delta_W$ is the Poincar\'{e} dual of $W$. (This corresponds to a non-trivial first Chern class of the bundle over $M_5 \! \setminus \! W$ preventing the extension over all of $M_5$~\cite{Witten:2009b}.) In particular, this singularity of $\cF$ prevents the extension of the equations of motion (\ref{eqn:YangMillsEquationOfMotion}) for $F$ to all of $M_5$, which can be easily seen since the LHS is not closed. The required modification is derived from (\ref{eqn:CurvatureSingularity}) as
\beq
\label{eqn:YangMillsEquationOfMotionModified}
-d\left( \frac{1}{R} \ast_g F \right) + \cF \wedge F = \delta_W \wedge J \,,
\eeq
where $J$ induces a one-form on $W$ through the pull-back of the canonical injection $W \hookrightarrow M_5$ which satisfies
\beq
\label{eqn:WZWEquationOfMotion}
dJ = - c_{\cF} F_{|W} \,.
\eeq
Since the field strength $F$ only determines $dJ$ the current represents additional degrees of freedom localized at $W$, whose dynamics is governed by the equation of motion (\ref{eqn:WZWEquationOfMotion}) for the gauged WZW model~\cite{Witten:1992} in agreement with the anomaly cancellation argument~\cite{Witten:2009b}. 

In order to verify that $J$ indeed contains additional degrees of freedom localized at the singularity we can consider the amount of initial value data that we must supply to be able to solve the system of differential equations describing the complete theory: For a regular fibration (or away from the singularity) we have the equations of motion derived in~\cite{Linander:2012} for the gauge field $F$, which together with the scalar and spinor fields comprise eight bosonic and eight fermionic degrees of freedom. The corresponding initial value data are required also in the singular case, but the additional equation (\ref{eqn:WZWEquationOfMotion}) on $W$ requires additional initial value data to be specified to solve for the current $J$.

\section{The Taub-NUT space}
The topic of this note being the single-centered Taub-NUT space we will now review its definition and some of its properties. The Taub-NUT is a hyper-K\"{a}hler manifold with metric
\beq
\label{eqn:TaubNutMetric}
ds_{\rm TN}^2 = U \delta_{ij} dx^i dx^j + \frac{1}{U} (d \varphi + \theta_i dx^i)^2 \,
\eeq
where $x^i$, with $i=1,2,3$, parametrize $\RR^3$ with the flat metric $\delta_{ij}$, $U$ is a function on $\RR^3$ given by
\beq
U = \frac{1}{\lambda^2} + \frac{1}{r} \,,
\eeq
$r = |\vec{x}|$ denotes the distance to the origin of $\RR^3$ and $dU = \ast_{\delta} d \theta$. (We denote by $\ast_{\delta}$ the Hodge dual on $(\RR^3,\delta)$.) The metric (\ref{eqn:TaubNutMetric}) has a $\U(1)$ isometry\footnote{In fact, (\ref{eqn:TaubNutMetric}) with an arbitrary function $U$ gives the most general form of a hyper-K\"{a}hler metric with a $\U(1)$ symmetry~\cite{Lindstrom:1983}.}, preserving the hyper-K\"{a}hler structure, corresponding to rotations of the compact $S^1$ parametrized by $\varphi$. Consequently, $\theta = \theta_i dx^i$ is interpreted as the connection one-form of a $\U(1)$ bundle over $\RR^3$ (away from singularities of $U$) and the corresponding field strength is denoted $\cF = d\theta$. The constant $\lambda$ gives the radius of the $\U(1)$ orbits at infinity\footnote{We adopt the notation of~\cite{Witten:2009a} in which the periodicity of the compact coordinate $\varphi$ is $4\pi$ in contrast to the case in~\cite{Linander:2012}. None of the considerations in the present paper depend on this convention.}. More specifically, connecting to the expressions in the previous section, the $S^1$ radius is $R = U^{-1/2}$ at a generic point in $\RR^3$. Thus, at the origin the fibre radius vanishes and the description of $TN$ as a $\U(1)$ bundle breaks down.

In order to describe the manifold $M_6 = \RR^{1,1} \times TN$ we decompose the index $\mu = (\sigma^a,x^i)$, where $\sigma^a$ are light-cone coordinates on $\RR^{1,1}$ in which the metric is $\eta_{+-} = 1$. The direct product metric implies that $\theta_{\mu}$ and $\cF_{\mu \nu}$ only have non-vanishing components in the $x^i$ directions so that we can identify (away from the origin in $\RR^3$) the Taub-NUT $\U(1)$ connection with the connection of the bundle over $M_5 \! \setminus \! W$. From the relation $\cF = \ast_{\delta} dU$ we can then explicitly compute the obstruction to extending the $\U(1)$ bundle to all of $M_5$
\beq
d \cF = -4 \pi \delta_{W}\,.
\eeq

On $TN$ there exists~\cite{Witten:2009a,Pope:1978,Eguchi:1979} a unique anti-self-dual harmonic two-form $\Omega$, expressible as $\Omega = d \Lambda$ with
\beq
\Lambda = \frac{1}{\lambda^2} U^{-1} (d\varphi+\theta_i dx^i) \,.
\eeq
It can be shown that $\Omega$ is $L^2$-normalizable on $TN$ (although its cohomology is non-trivial since $\Lambda$ is not normalizable) and supported near the singularity at $|\vec{x}|=0$. Using $\Omega$ we can then engineer a solution to the equations of motion ($dH=0$ and $H=\ast_G H$) for the three-form field strength of the tensor multiplet on $M_6$ as
\beq
\label{eqn:OmegaSolution}
H = h_+(\sigma^+)d\sigma^+ \wedge \Omega \,,
\eeq
where $h_+(\sigma^+)$ is an arbitrary function. (The present geometry can be viewed as a special case of a multi-centered Taub-NUT geometry, in which there exists a number of anti-self-dual closed two-forms $\Omega_I$ and a solution to the self-duality equation of motion is obtained by a linear combination generalizing (\ref{eqn:OmegaSolution})~\cite{Lambert:2012c}.)
From the properties of $\Omega$ it follows that this solution is well-defined on all of $M_6$ and in particular localized near the singularity $W$. Away from $W$ we can use the decomposition (\ref{eqn:DecompositionH}) to extract the field strength $F$
\beq
F_{i+} = - \frac{h_+(\sigma^+) \lambda^2}{r(r+\lambda^2)^2} x_i \,\,\, , \,\,\, F_{ij} = 0 \,\,\, , \,\,\, F_{i-} = 0 \,\,\, , \,\,\, F_{+-} = 0 \,.
\eeq
Extending the solution to all of $M_5$ (i.e. including the origin $|\vec{x}|=0$) we can compute the current $J$ from (\ref{eqn:YangMillsEquationOfMotionModified}) as
\beq
\label{eqn:TaubNUTCurrent}
J = - 4\pi h_+(\sigma^+) d\sigma^+ \,.
\eeq

We note that in agreement with $J$ being the current of a gauged WZW model it contains only left-moving excitations, which is precisely what is required to cancel the anomalous gauge transformation of the gauge field.

\section{Spherically symmetric solutions}
We will now proceed to consider solving the equations of motion for the Yang-Mills field strength $F_{\mu \nu}$ in the case of $M_6 = \RR^{1,1} \times TN$. We will assume spherical symmetry in $\RR^3$ which restricts the form of the components of $F_{\mu \nu}$ to
\bea
\label{eqn:SphericalSymmetryAssumption1}
F_{i j} & = & \varepsilon_{i j k} x^k f(\sigma,r) \\
\label{eqn:SphericalSymmetryAssumption2}
F_{i a} & = & x_i f_{a}(\sigma,r) \\
\label{eqn:SphericalSymmetryAssumption3}
F_{a b} & = & f_{a b}(\sigma,r) \,,
\eea
for some functions that depend arbitrarily on $\sigma^{\alpha}$ but only on the distance $r = |\vec{x}|$ to the origin of $\RR^3$. We will consider the solution to the equations of motion (\ref{eqn:YangMillsEquationOfMotion}) away from the singularity at $|\vec{x}|=0$. In doing so, we require that the  components $F_{\mu \nu}$ are regular at $W$. In principle, the coupling to the degrees of freedom localized at the singularity $W$ can then be computed as above by extending the solutions over $|\vec{x}|=0$.

Considering first the Bianchi identity $dF = 0$ we extract the differential equation
\beq
f + \frac{1}{3} r \d_r f = 0 \,,
\eeq
which implies $f \sim r^{-3}$. The divergence as $r \to 0$ and required regularity of $F$ hence implies that $F_{ij}$ must be identically vanishing. Inserting this into the remaining component equations of $dF = 0$ we find the condition
\beq
\d_r f_{+ -} = r (\d_+ f_- - \d_- f_+) \,.
\eeq
Moving on to the equation of motion $-d(U^{1/2} \ast_g F) + \cF \wedge F = 0$ we find the remaining differential equations determining the functions $f$, $f_{a}$ and $f_{ab}$:
\beq
\label{eqn:PDE1}
-\left( \frac{1}{\lambda^2} + \frac{1}{r} \right)^2 \d_+ f_{+-} + \left( 3 \frac{1}{\lambda^2} + \frac{1}{r} \right)f_+ + r\left( \frac{1}{\lambda^2} + \frac{1}{r} \right) \d_rf_+ = 0
\eeq
\beq
\label{eqn:PDE2}
\left( \frac{1}{\lambda^2} + \frac{1}{r} \right)^2 \d_- f_{+-} + 3 \left( \frac{1}{\lambda^2} + \frac{1}{r} \right)f_- + r\left( \frac{1}{\lambda^2} + \frac{1}{r} \right) \d_r f_- = 0
\eeq
\beq
\label{eqn:PDE3}
f_{+ -} = -r^3 \left( \frac{1}{\lambda^2} + \frac{1}{r} \right) \left[ \d_+ f_- + \d_- f_+ \right] \,.
\eeq
In particular, we can use (\ref{eqn:PDE3}) to transform (\ref{eqn:PDE1}) and (\ref{eqn:PDE2}) into a system of equations for $f_+$ and $f_-$.

Next, we consider performing a Fourier expansion in the directions normal to $TN$ of the component $F_{+-}$ appearing in the WZW equation of motion in terms of eigenfunctions to the wave-operator on $\RR^{1,1}$. Since the theory on $M_5$ is linear we can consider the components independently. We therefore introduce the two-dimensional mass eigenvalue as
\beq
\label{eqn:MassEigenvalue}
\d_+ \d_- F_{+-} = -\frac{1}{2} m^2 F_{+-} \,.
\eeq
Taking further derivatives of the original equations and inserting the mass eigenvalue we obtain
\beq
\label{eqn:PDE4}
r\left(\frac{1}{\lambda^2} + \frac{1}{r} \right) \d_r \d_+ f_- + 3 \left(\frac{1}{\lambda^2} + \frac{1}{r} \right) \d_+ f_- + \frac{1}{2} m^2 r^3 \left( \frac{1}{\lambda^2} + \frac{1}{r} \right)^3 (\d_+ f_- + \d_- f_+ ) = 0
\eeq
\beq
\label{eqn:PDE5}
r\left(\frac{1}{\lambda^2} + \frac{1}{r} \right) \d_r \d_- f_+ + \left(3 \frac{1}{\lambda^2} + \frac{1}{r} \right) \d_- f_+ - \frac{1}{2} m^2 r^3 \left( \frac{1}{\lambda^2} + \frac{1}{r} \right)^3 (\d_+ f_- + \d_- f_+ ) = 0
\eeq
which, treated as a coupled system of first order equations in $r$, can be solved for the derivative quantities $\d_+ f_-$ and $\d_- f_+$.

When $m=0$ the system is greatly simplified by the decoupling of the equations (\ref{eqn:PDE4}) and (\ref{eqn:PDE5}) for $\d_+f_-$ and $\d_- f_+$, and we will only consider this case in detail here. The decoupled system of equations is solved by
\bea
\label{eqn:PDE6}
\d_- f_+ & = & \tilde{c}_+(\sigma) \frac{1}{r(r+\lambda^2)^2}\\
\label{eqn:PDE7}
\d_+ f_- & = & \tilde{c}_-(\sigma) \frac{1}{r^3} \,.
\eea
In order to have $F_{+-}$ regular we must impose $\tilde{c}_-(\sigma)=0$, which implies that
\beq
f_{+-} = -\frac{1}{\lambda^2}\frac{\tilde{c}_+ r}{(r+\lambda^2)} 
\eeq
Furthermore, the eigenvalue equation (\ref{eqn:MassEigenvalue}) implies that $\tilde{c}_+ = c_+(\sigma^+) + c_-(\sigma^-)$.

Integrating (\ref{eqn:PDE6}) and (\ref{eqn:PDE7}) with respect to $\sigma^+$ and $\sigma^-$ respectively we obtain expressions for $f_+$ and $f_-$. In order to determine the dependence on $r$ in the constants of integration, which we will denote $d_+(\sigma^+,r)$ and $d_-(\sigma^-,r)$, we reinsert the expression into (\ref{eqn:PDE1}) and (\ref{eqn:PDE2}) obtaining the equations
\beq
\label{eqn:PDEConstantOfIntegration1}
\left( 3 \frac{1}{\lambda^2} + \frac{1}{r} \right)d_+ + r\left( \frac{1}{\lambda^2} + \frac{1}{r} \right) \d_r d_+ = -\frac{1}{\lambda^4}\left( \frac{1}{\lambda^2} + \frac{1}{r} \right) \d_+ c_+
\eeq
\beq
\label{eqn:PDEConstantOfIntegration2}
3 \left( \frac{1}{\lambda^2} + \frac{1}{r} \right)d_- + r\left( \frac{1}{\lambda^2} + \frac{1}{r} \right) \d_r d_- = \frac{1}{\lambda^4}\left( \frac{1}{\lambda^2} + \frac{1}{r} \right) \d_- c_- \,.
\eeq
We discard, as before, solutions that produce divergences in the field strength $F_{\mu \nu}$ as $r \to 0$. Furthermore, we must also require that the action is finite for the solutions which leaves the unique solution
\beq
f_+ = \frac{\tilde{h}_+(\sigma^+)}{r(r+\lambda^2)^2} \,\,\, , \,\,\, f_- = 0 \,\,\, , \,\,\, f_{+-} = 0 \,,
\eeq
for some arbitrary function $\tilde{h}_+(\sigma^+)$. Up to a multiplicative constant this reproduces the field strength obtained from the six-dimensional solution $H = h_+(\sigma^+) d\sigma^+ \wedge \Omega$, constructed above using the anti-self-dual two-form on $TN$.

\section{Summary and conclusion}
In this note we consider the dimensional reduction of the $(2,0)$ tensor multiplet on singular circle fibrations. In particular, we consider a manifold $M_6$ with a $\U(1)$ action with a codimension four fixed point locus $W$, and the coupling through the current $J$ appearing in the modified gauge field equations of motion to a WZW model localized on the singularity $W$. As an example we study $M_6= \RR^{1,1} \times TN$, whose description as a $\U(1)$ fibration breaks down at the origin $|\vec{x}| = 0$ of the $\RR^3$ underlying the $TN$.

We also consider the the solution to the gauge field equation of motion away from $|\vec{x}|=0$ for a spherically symmetric ansatz. For the massless mode of $F_{+-}$ on $W$ we find that the solution is given by the the reduction of the solution $H=h_+(\sigma^+) d\sigma^+ \wedge \Omega$ (where $\Omega$ is the unique anti-self-dual closed two-form on $TN$) to the tensor field equations of motion in the six-dimensional $(2,0)$ theory. The regular part of the solution for $m \neq 0$, which we do not consider in detail, can be expressed in terms of generalized Laguerre functions $L_{n}^{\alpha}(r)$ and an exponential factor which localizes the solutions on $W$.

In addition to providing a concrete and relatively simple example of a singular circle fibration, a motivation for considering $TN$ is that a consequence of its hyper-K\"{a}hler structure (and the direct product metric on $M_6$) is the existence of covariantly constant spinors. The requirement for unbroken supersymmetry in six dimensions is the existence of transformation parameters satisfying the conformal Killing spinor equation on $M_6$. Since covariantly constant spinors certainly satisfy this requirement the geometry $M_6 = \RR^{1,1} \times TN$ permits non-trivial supersymmetry transformations of the tensor multiplet fields. In line with the present note it would be interesting to investigate the modification, induced by the singularity of the fibration, to the supersymmetry of the low energy effective gauge theory obtained in the dimensional reduction.
\vspace*{0.3cm}

The author wishes to acknowledge M{\aa}ns Henningson and Hampus Linander for fruitful discussions and valuable advice, and Neil Lambert for bringing results regarding the multi-center Taub-NUT case to his attention. This work was supported by grants from the Swedish Research Council and the G\"{o}ran Gustafsson Foundation.


\end{document}